\documentclass{iaus}
\usepackage[latin1]{inputenc}
\usepackage{graphicx}
\usepackage{aas_macros}

\title[SPH chemical evolution of bulges] 
{SPH simulations of the chemical evolution of bulges}

\author[F.J. Martínez-Serrano, R. Domínguez-Tenreiro, M. Mollá]
{F.J. Martínez-Serrano$^{1}$, R. Domínguez-Tenreiro$^{2}$, M.
Mollá$^{3}$}

\affiliation{$^{1}$Depto. de Física y A.C., Univ. Miguel Hernández,
03206 Elche, Alicante, Spain
\\[\affilskip]
 $^{2}$Depto. de Física Teórica, Univ. Autónoma de Madrid,
28040 Cantoblanco, Madrid, Spain\\[\affilskip]
$^{3}$Depto. de Investigación Básica, C.I.E.M.A.T., Avda. Complutense 22, 28040
Madrid, Spain}

\pubyear{2007}
\volume{IAUS245}  
\pagerange{119--126}
\date{?? and in revised form ??}
\jname{Proceedings Title IAU Symposium}
\editors{A.C. Editor, B.D. Editor \& C.E. Editor, eds.}

\begin{document}

\maketitle

\begin{abstract}
We have implemented a chemical evolution model on the parallel
AP3M+SPH DEVA code which we use to perform high resolution simulations
of spiral galaxy formation. It includes feedback by SNII and SNIa
using the Q$_{ij}$ matrix formalism. We also include a diffusion
mechanism that spreads newly introduced metals.  The gas cooling rate
depends on its specific composition.  We study the stellar populations
of the resulting bulges finding a potential scenario where they seem
to be composed of two populations: an old, metal poor,
$\alpha$-enriched population, formed in a multiclump scenario at the
beginning of the simulation and a younger one, formed by slow
accretion of satellites or gas, possibly from the disk due to instabilities.

\end{abstract}

\firstsection 

\section{Method}

For the chemical enrichment, we adapt the Q$_{ij}$ formalism
\cite{1971ApJ...170..409T} to the SPH method, thus accounting for the
nucleosynthetic sources of each element created in star particles. The Q$_{ij}$
matrices explicitely depend on the total metallicity Z. The star particles
create new elements at each timestep and release them to the closest gas
neighbor. A SPH diffusion algorithm then accounts for the redistribution of
elements in the gas due to interstellar turbulence.

For the cooling rate, we use the Mappings III code \cite{1993ApJS...88..253S}
to compute the cooling function of $\sim 10^5$ different compositions, all of
them extracted from full cosmological simulations. We then apply a dimension
reduction regression \cite{1991JASA...86..316L} algorithm to make the cooling
functions dependent on a single parameter.  We thus keep the dependence on 16
elemental abundances without high computational cost (see Mart\'{\i}nez-Serrano
et al., in preparation for details on the chemical evolution and coling models).

\section{Results: bulges from cosmological simulations}

We have run three high resolution cosmological simulations to obtain three
spiral galaxies using the multimass technique. All three objects have been
extracted from a single low-resolution (2x64$^3$) full $\Lambda$CDM
cosmological simulation with parameters $(\Omega_M, \Omega_\Lambda, \Omega_b,
h, \sigma_8, L) = (0.3, 0.7, 0.04, 0.7, 1, 10 \textrm{ Mpc})$ that included all
the processes described above. The baryonic mass resolution for the
resimulation of each of the objects is $(m_{b1},m_{b2},m_{b3}) =
(29.24,3.51,3.51)\, 10^5$.  These objects have good dynamical properties (see
contributed talk by Domínguez-Tenreiro in this volume), as well as colors and
sizes.

The SFRH in the bulge have two clearly differenciated stages. The
first (namely old) stage is a multiclump-dominated phase with a high
SFR and the second (namely young) stage to be more quiescent with star
formation arising mainly due to gas infall. We find that the separation
between both phases is at $t_c \simeq 0.32 t_0$ or $z_c \simeq 1.5$.

\begin{figure*}
\begin{center}
\begin{tabular}{ccc}
 (a) & (b) & (c)\\
 \includegraphics[angle=-90,width=0.33\textwidth]{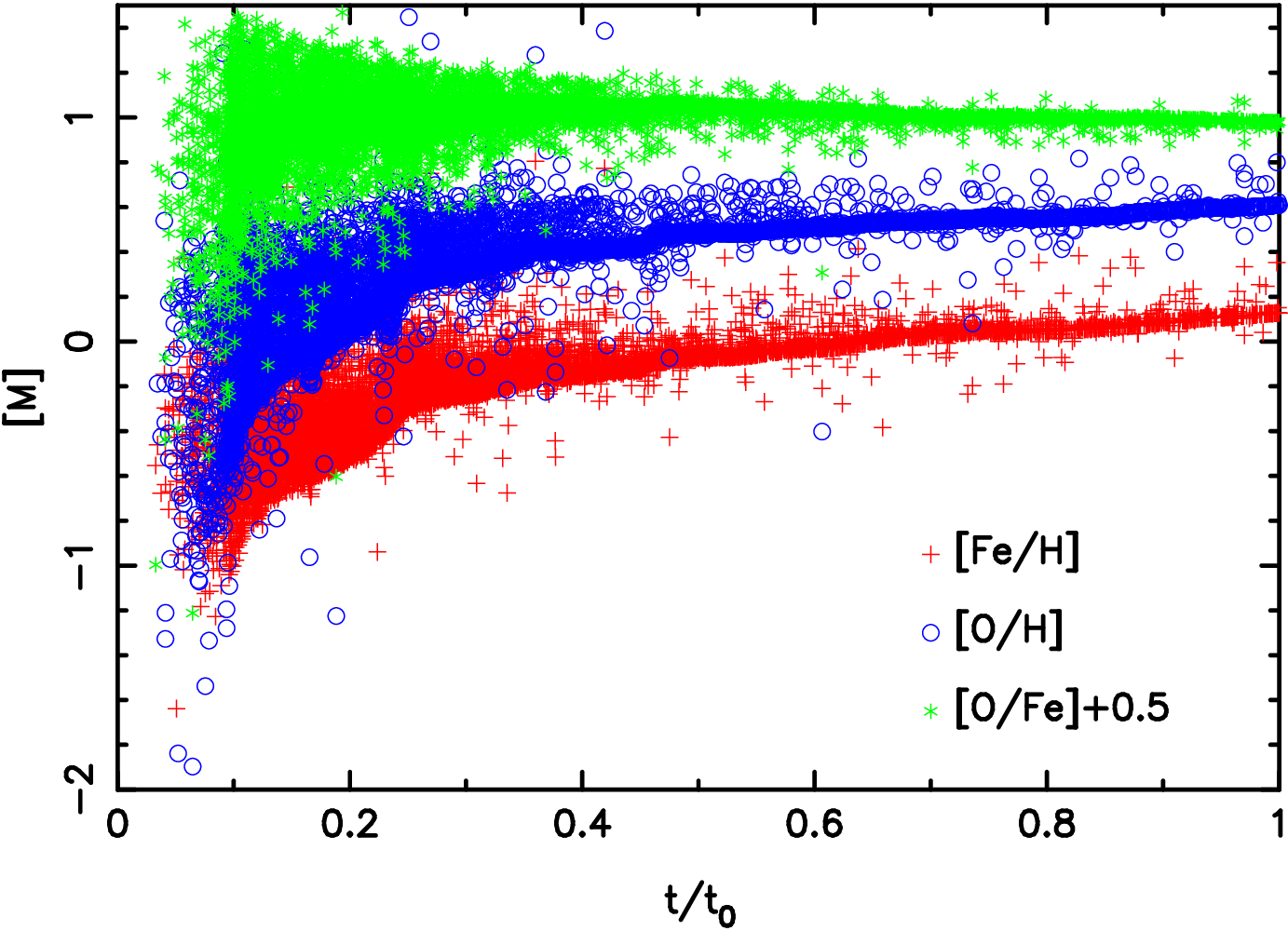} &
 \includegraphics[angle=-90,width=0.33\textwidth]{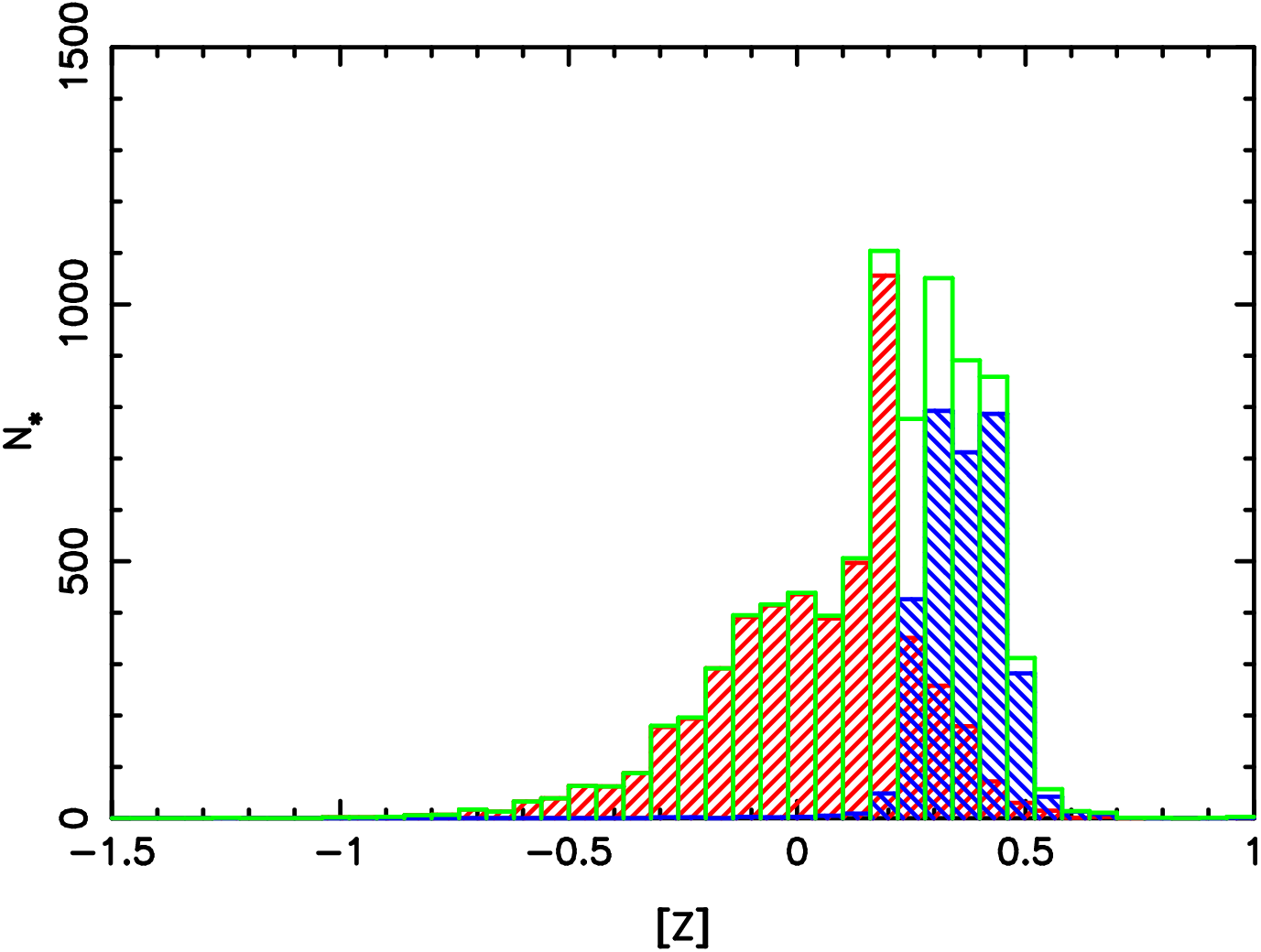} &
 \includegraphics[angle=-90,width=0.33\textwidth]{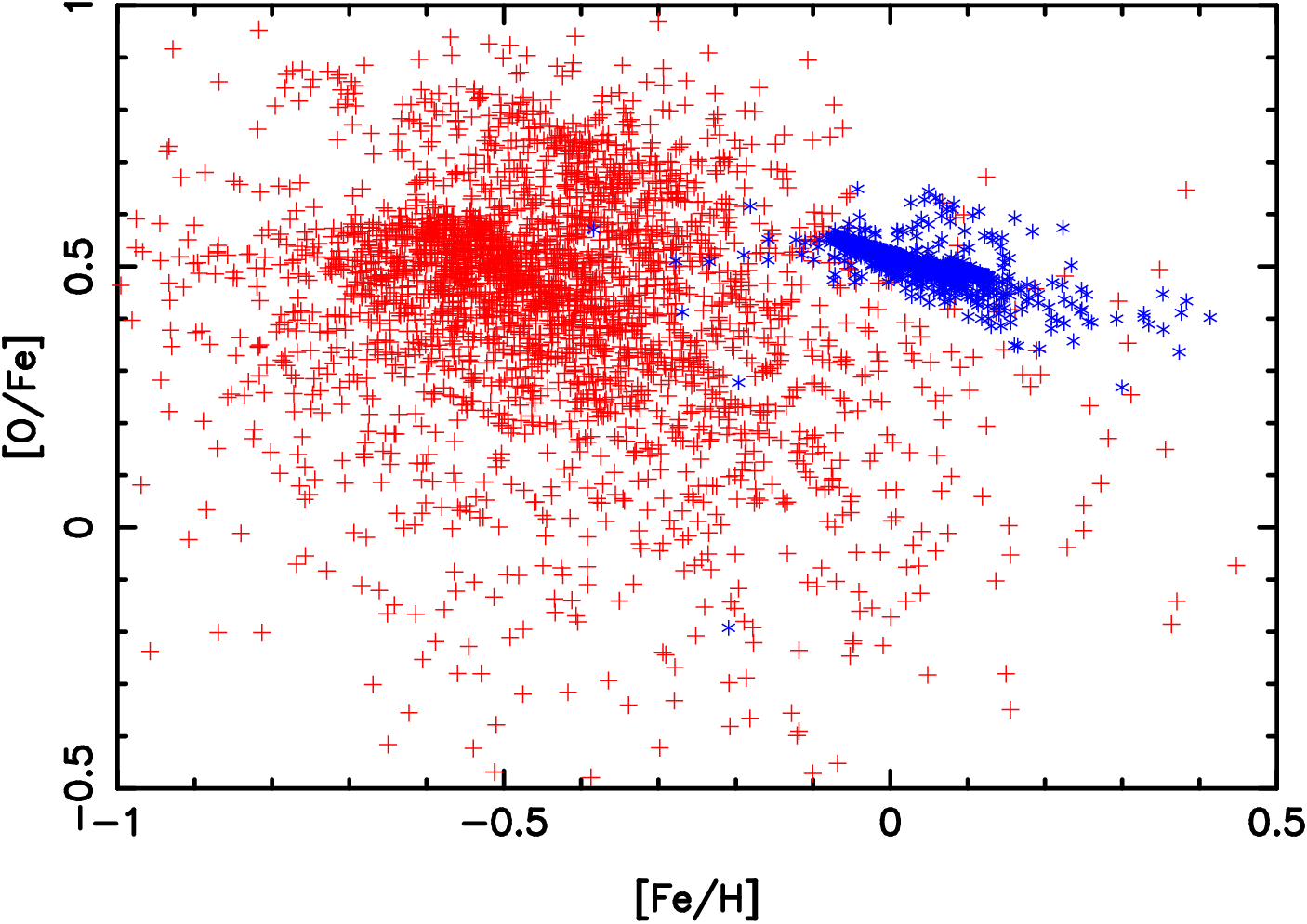} \\
\end{tabular}

\caption{\label{fig}}
\end{center}
\end{figure*}

The age-metallicity relation for the present bulge stars (Fig. \ref{fig}-a)
shows this split between both phases as a change in the slope of the
metallicity at $t_c$.  [O/H] grows faster than [Fe/H], producing a higher
[O/Fe] relation for stars formed at earlier times.

The metallicity distribution function (MDF) (Fig. \ref{fig}-b) also shows a
clear bimodality, with the old stars having a greater spread in total
metallicity and typically lower values, such as those from elliptical galaxies.
Meanwhile the young component, mostly formed from gas in the disk (see talk by
Domínguez-Tenreiro), has less dispersion and a higher mean value. The
composition of both distributions produces the typical shape observed in the
bulge of M31, peaking around [Fe/H]=0, with a steep decline at higher
metallicities and a elongated tail to lower metallicities
\cite{2005AJ....130.1627S}. The observed skew in the distribution is explained
in our simulations as a consequence of the bimodality of the stellar
populations: both MDFs are almost symmetrical when considered separately, but
their ensemble has a significant skew.

The alpha-metallicity relation (Fig.~\ref{fig}-c) also shows two different
shapes for both components, with the young population (stars) having an slope
of $\sim -0.25$, consistent with observations of disk and bulge
\cite{2007ApJ...661.1152F} and also with values predicted by standard chemical
evolution models \cite{2000MNRAS.316..345M}.

\section{Conclusions} We have performed three high resolution simulations of
disk formation using our new P-DEVA code. The three bulges obtained have
similar properties, consistent with a formation scenario in two phases: a fast,
multiclump dominated one and a secular evolution phase with sporadic mergers.
The shift between both phases appears at $z_c \sim 1.5$. Two stellar
populations with different chemical properties appear accordingly. At $z_c$,
the old component is already in place, while the young one is still a disk
component and will not be completely in the bulge until much lower redshifts
($z \sim 0.2$).

\end{document}